\documentclass[aip,reprint,amsmath,amssymb]{revtex4-1}
\usepackage{graphicx}
\usepackage{dcolumn}
\usepackage{bm}
\usepackage[mathlines]{lineno}

\usepackage{xcolor}
\usepackage[utf8]{inputenc}
\usepackage[T1]{fontenc}
\usepackage{times}
\usepackage{etoolbox}
\usepackage{stackrel} 
\usepackage[unicode=true,pdfusetitle,
bookmarks=true,bookmarksnumbered=false,bookmarksopen=false, breaklinks=true, ,backref=false,colorlinks=true]{hyperref}
\hypersetup{citecolor=blue}

\begin{document}


\title{A Principled Basis for Nonequilibrium Network Flows}
\author{Ying-Jen Yang}
\email{ying-jen.yang@stonybrook.edu}
 \affiliation{Laufer Center for Physical and Quantitative Biology, Stony Brook University, NY, USA}
\author{Ken A. Dill}
\email{dill@laufercenter.org}
\affiliation{Laufer Center for Physical and Quantitative Biology, Stony Brook University, NY, USA}
\affiliation{ 
Department of Physics, Stony Brook University, NY, USA}%
\affiliation{ 
Department of Chemistry, Stony Brook University, NY, USA}

\date{\today}

\begin{abstract}
The great power of EQuilibrium (EQ) statistical physics comes from its principled foundations: its First Law (conservation), Second Law (variational tendency principle), and its Legendre Transforms from observables $(U, V, N)$ to their driving forces $(T, p, \mu)$. Here, we generalize this structure to Non-EQuilibria (NEQ) in \textit{Caliber Force Theory} (CFT), replacing state entropies with path entropies; and $(U, V, N)$ with dynamic observables (node probabilities, edge traffics, and cycle fluxes).  CFT derives dynamical forces and a complete set of conjugate relations: (i) It yields generalized Maxwell-Onsager relations, applicable far from equilibrium; (ii) It constructs dynamical models from mixed force-observable constraints; and (iii) It reveals new relationships---including an ``equal-traffic'' rule for optimizing molecular motors, and a ``third Kirchhoff's law'' of stochastic transport---and can resolve some dynamical paradoxes. \\
\end{abstract}

\maketitle

\subsection*{Caliber Force Theory: Two Laws of Nonequilibria}
Nonequilibrium (NEQ) statistical physics has not yet had the same level of solid and general grounding in foundational principles as equilibrium (EQ) statistical physics has had.\cite{touchette_large_2009}  In EQ physics,\cite{huang_statistical_1987,landau_statistical_1980} a phase space is represented by a complete set of observables such as $(U, V, N)$ and is subject to conservation (First Law) and an entropic tendency (Second Law).  Legendre Transforms then give \textit{forces,} such as $(1/T, ~p/T, ~-\mu/T)$, as well as Maxwell Relations and near-EQ fluctuation-response relations. Examples of the many practical manifestations of EQ physics are: (1) Computing forces is essential in modeling molecules; their net balance identifies stable states and dominant populations.  (2) Maxwell Relations give conversions to compute experimentally realizable properties from theoretical model quantities such as $S(U)$.  In addition, early theories of NEQ took their cue from EQ, starting from $(U, V, N)$ or $(T,P,\mu)$ and the First and Second Laws of EQ, but in so doing such models were limited to near or local-EQ processes, \textit{e.g.} requiring baths to be fast-equilibrating for well-defined $T$ and $\mu$.\cite{pachter_nonequilibrium_2023,pachter_entropy_2024} \\

We give here a general framework of principles for NEQ statistical physics, which we call \textit{Caliber Force Theory} (CFT).  In important ways, it resembles the framework of EQ: it entails a conservation law; a variational tendency principle; Legendre Transforms that give forces; Maxwell-like relations among forces and observables; and fluctuation-response relations. However, for NEQ, we require two main differences relative to EQ theory.  First, the Second-Law-like variational principle for NEQ is Maximum Caliber (Max Cal),\cite{jaynes_minimum_1980,evans_rules_2004,presse_principles_2013,davis_probabilistic_2018} maximizing an entropy \textit{over pathways}, not Maximum Entropy, maximizing an entropy \textit{over states}.  Second, the complete set of observables is not the $(U, V, N)$ quantities of EQ theory, but rather $(\pi, \tau, J)$, where $\pi$s are state probabilities at the nodes; $\tau$s are the total flux activity (``traffic''\cite{maes_canonical_2008,maes_frenesy_2020}) along the edges, and $J$s are net fluxes around the cycles.\cite{schnakenberg_network_1976,hill_free_1989} A key step stone here is the construction of a non-degenerate basis for these observables, forming the proper coordinate system required for their Legendre transforms. 

\subsection*{Two Laws Yield a Complete Conjugate Structure}
\begin{figure}
\begin{centering}
\includegraphics[width=0.83\columnwidth]{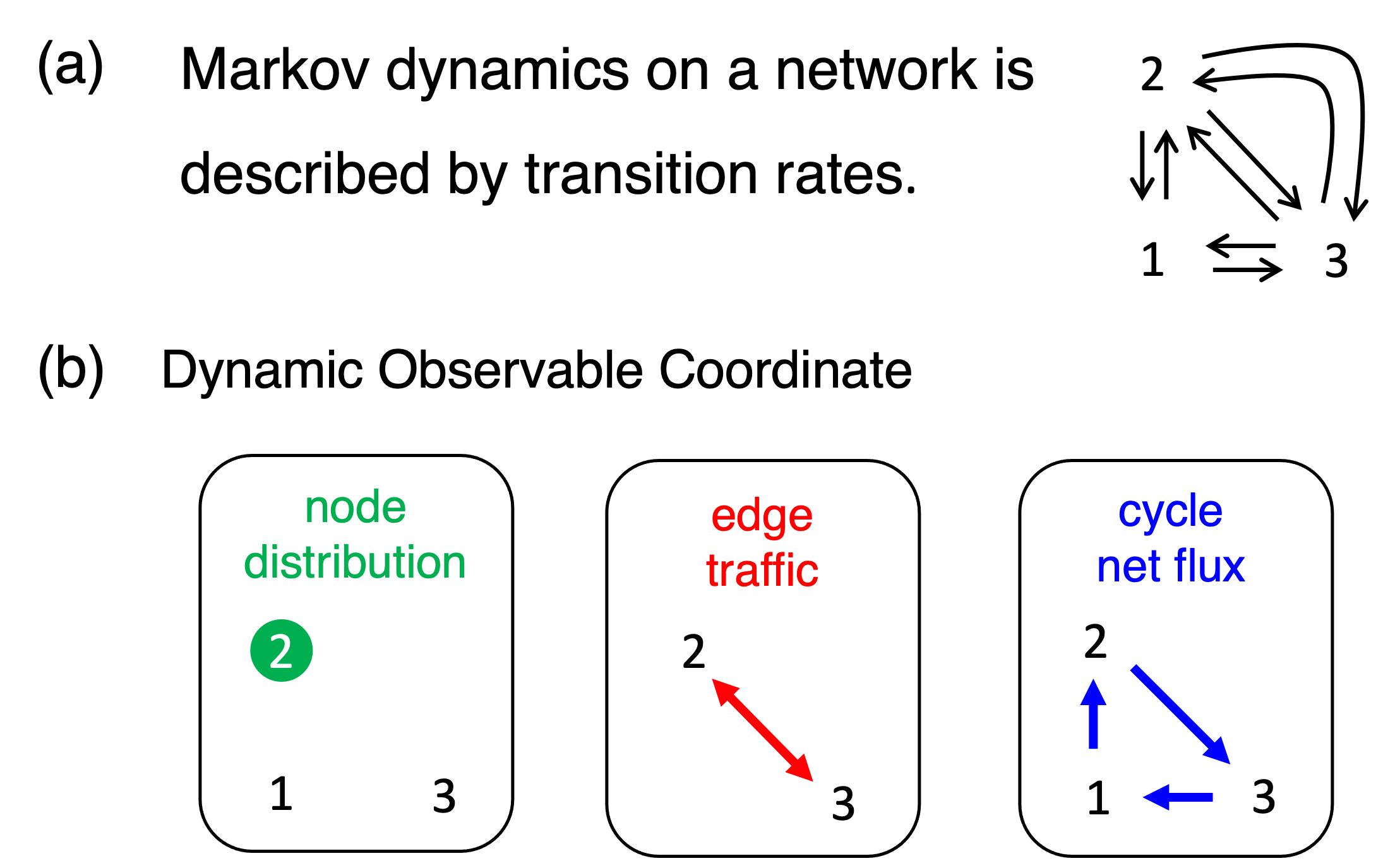}
\par\end{centering}
\caption{ 
\textbf{The three observables: node populations, edge traffic and cycle fluxes}:  (a) Simple example network of nodes and edges with forward and reverse Markov transition rates.  
(b) (Left) Each node has a steady-state probability $\pi_{\text{node}}$.  (Middle) Each edge has \textit{traffic} $\tau_{\text{edge}}$, the sum of forward plus back fluxes.  (Right) Net flows at the steady state are circular $J_{\text{cycle}}$.  
\label{fig: Parsing} }
\end{figure}

We establish thermodynamic-like coordinate systems for NEQ steady states based on two foundational principles (see Method A for why steady state). 
The \textbf{First Law is a principle of completeness derived from the conservation of probability}. It establishes that a steady-state Markov jump process on a network (Fig. \ref{fig: Parsing}a) can be characterized by a complete and non-degenerate set of observables:  $\textbf{x} = (\pi_{\text{node}}, \tau_{\text{edge}}, J_{\text{cycle}})$ (Fig. \ref{fig: Parsing}b). The coordinate must be non-degenerate: the vector of node probabilities, $\pi_{\text{node}}$, omits one probability (from a chosen reference node, $m$) to handle normalization, while the net fluxes are described by a linearly-independent basis of fundamental cycle fluxes, $J_{\text{cycle}}$, to handle steady-state flux conservation.\cite{schnakenberg_network_1976,hill_free_1989}  This basis parameterizes the underlying transition rates $k_{ij}$ (from node $i$ to $j$):
\begin{equation} \label{eq: k from x}
k_{ij}=\frac{\tau_{\text{edge},ij}+J_{ij}(J_{\text{cycle},c})}{2\pi_{\text{node},i}}
\end{equation}
where edge net fluxes $J_{ij}$ is spanned by the cycle fluxes $J_c$. See Method B for the steps to construct $\textbf{x}$. The dimension parsing here is similar to the landscape-flux decomposition in Markov diffusion processes.  \cite{graham_covariant_1977,ao_potential_2004,wang_potential_2008,yang_potentials_2021}
\\

The \textbf{Second Law is the Principle of Maximum Caliber (Max Cal)},\cite{jaynes_minimum_1980, evans_rules_2004,presse_principles_2013,davis_probabilistic_2018} which identifies the \textit{path entropy rate} as the potential of NEQ. Relative to a reference ``doldrum'' process with unit transition rates $u_{ij}=1$, it is given by \begin{subequations} \label{eq: path entropy}
    \begin{align}
    \mathfrak{s}_{\text{path}}(\textbf{x}) &= \lim_{L\rightarrow \infty} \frac{1}{L}\left[-\sum_{\text{path }\omega_L} P_k(\omega_L) \ln \frac{P_k(\omega_L)}{P_{u}(\omega_L)}\right]\\
    &=\sum_{i \neq j} p_{ij} (\textbf{x}) - \pi_i (\textbf{x})- p_{ij} (\textbf{x})\ln \frac{p_{ij}(\textbf{x})}{\pi_i(\textbf{x})}.\label{eq: spath wrt doldrum}
    \end{align}
\end{subequations} 
where $P_k$ is the path probability of a process with transition rates $k_{ij}$ and $p_{ij}(\textbf{x})=[\tau_{ij}+J_{ij}(J_c)]/2$ is the probability edge flux as a function of traffic and cycle fluxes.\\

These two laws establish a pair of conjugate thermodynamic-like potentials. The path entropy, $\mathfrak{s}_{\text{path}}(\textbf{x})$, serves as the potential in the observable coordinate $\textbf{x}$. 
We define the \textbf{conjugate forces}, $\boldsymbol{\mathfrak{F}}$, as its derivatives. CFT's Two-Law procedure yields physically interpretable expressions for the forces as affinities for node dwelling, edge exchange, and directed cycle completion (Method C):
\begin{align}\label{eqs: force-expression-in-terms-of-the-rates} 
\mathfrak{F}_{\text{node},n} & =-\frac{\partial \mathfrak{s}_{\text{path}}}{\partial \pi_n}=\sum_{i(\neq m)} (k_{mi}-1)-\sum_{j(\neq n)} (k_{nj}-1),\nonumber\\
\mathfrak{F}_{\text{edge},ij} &=-\frac{\partial \mathfrak{s}_{\text{path}}}{\partial \tau_{ij}} =  \frac{1}{2} \ln {k_{ij}  k_{ji} },\nonumber\\
\mathfrak{F}_{\text{cycle},c} &=-\frac{\partial \mathfrak{s}_{\text{path}}}{\partial J_c} =  \frac{1}{2} \ln {\frac{k_{i_{0}i_{1}} k_{i_{1}i_{2}}\cdots k_{i_{\sigma} i_{0}}}{k_{i_{0}i_{\sigma}} k_{i_{\sigma} i_{\sigma-1}} \cdots k_{i_{1}i_{0}}}}
\end{align}
where 
$i_0i_1i_2...i_\sigma i_0$ is the state sequence of the cycle $c$. 
This novel physical interpretability is what makes force constraints in our later analysis meaningful.
Importantly, this shows that the cycle force in CFT is exactly half the cycle affinity in Stochastic Thermodynamics (ST),\cite{schnakenberg_network_1976,hill_free_1989,yang_bivectorial_2021} demonstrating---for the first time---a long-sought variational-principled relationship between cycle affinity and cycle flux.\\

The second potential, \textbf{the Caliber} $\mathfrak{c}(\boldsymbol{\mathfrak{F}})$, is defined via a \textbf{Legendre transform}, creating a full duality between the force and observable coordinates:
\begin{equation} \label{eq: LT}
    \mathfrak{c}(\boldsymbol{\mathfrak{F}}) =  \boldsymbol{\mathfrak{F}}\cdot\textbf{x}+\mathfrak{s}_{\text{path}}(\textbf{x}).
\end{equation} 
In direct analogy to the \textit{free energy} in equilibrium physics, the Caliber is the (scaled) log-partition function over all paths $\omega_L$:
\begin{equation} 
\mathfrak{c}(\boldsymbol{\mathfrak{F}}) = \lim_{L\to\infty} \frac{1}{L} \ln \sum_{\omega_L} P_u(\omega_L) e^{\boldsymbol{\mathfrak{F}} \cdot \textbf{X}(\omega_L)}
\label{eq:caliber_def}
\end{equation}
where $\textbf{X}(\omega_L)$ is the vector of time-extensive counting path observables whose long-term average rates are the observable coordinate $\textbf{x}$ (see Method B for explicit expressions). Large Deviation Theory (LDT) offers a simple way to compute the Caliber (Method D):\cite{mullins_analysis_1959,touchette_large_2009,barato_formal_2015,chetrite_nonequilibrium_2015} it is the largest eigenvalue of ``the tilted matrix'', constructed by ``tilting'' the transition rate matrix of the doldrum process with the forces. \\

This dual-potential structure empowers our framework. It provides a complete one-to-one map between the observable and force coordinates: the path entropy transforms observables to forces (Eq. \ref{eqs: force-expression-in-terms-of-the-rates}), while the Caliber gives the reverse: \begin{equation}
    \label{eq: x = dc/dF}
    \textbf{x}(\boldsymbol{\mathfrak{F}})=\frac{\partial \mathfrak{c}}{\partial \boldsymbol{\mathfrak{F}}}.  
\end{equation}
This structure immediately yields a complete set of universal response laws. The second derivatives of the Caliber produce the \textbf{Fluctuation-Susceptibility Equalities}, which include the far-from-EQ generalizations of \textbf{Maxwell relations} at EQ and \textbf{Onsager reciprocality} near EQ:\cite{onsager_reciprocal_1931}\begin{subequations} \label{fluctuations}
    \begin{align}
        \frac{\partial x}{\partial \mathfrak{F}} &=\frac{\partial^2  \mathfrak{c}}{\partial \mathfrak{F}^2} (\boldsymbol{\mathfrak{F}}) =\lim_{L \rightarrow\infty} \frac{\text{Var}[X]}{L} >0; \label{eq: FRR conjugate}\\
 \frac{\partial x}{\partial \mathfrak{F}'}=\frac{\partial x'}{\partial \mathfrak{F}}&=\frac{\partial^2 \mathfrak{c}}{\partial \mathfrak{F} \partial \mathfrak{F}'}(\boldsymbol{\mathfrak{F}}) = \lim_{L\rightarrow\infty} \frac{\text{CoV}[X,X']}{L}.\label{eq: FRR covariance}
    \end{align}
\end{subequations}
where Var and CoV represent variance and covariance, $X,X'$ can be any two of the counting variables $\textbf{X}$,  $x,x'$ are their average rates, and $\mathfrak{F},\mathfrak{F}'$ are their conjugated forces.\\

A power of equilibrium physics is the ability to work with \textit{mixtures} of observables and forces, such as energy and temperature.  We have the same power here.  For a mixed coordinate $\boldsymbol{z}=(\boldsymbol{\mathfrak{F}}_1,\textbf{x}_2)$ obtained by switching out conjugate variables, partial Legendre transforms define the corresponding potential $\phi(\boldsymbol{z})$:
\begin{subequations} \label{eqs: phi}
\begin{align}
        \phi &= \boldsymbol{\mathfrak{F}}_1\cdot\textbf{x}_1 +\mathfrak{s}_{\text{path}}; \label{eq: phi and s}\\
        -\phi &= \boldsymbol{\mathfrak{F}}_2\cdot \textbf{x}_2-\mathfrak{c}. \label{eq: phi and c}
    \end{align}
\end{subequations} 
The remaining unknown variables are its derivatives:
\begin{equation} \label{eqs: computing x and F from d phi /dF and x}
\textbf{x}_1(\boldsymbol{z}) = \frac{\partial \phi}{\partial \boldsymbol{\mathfrak{F}}_1} ~~~;~~~
\boldsymbol{\mathfrak{F}}_2(\boldsymbol{z}) = -\frac{\partial \phi}{\partial \textbf{x}_2}.
\end{equation}
The underlying transition rates $k_{ij}$ can then be computed by combining Eq. \eqref{eqs: computing x and F from d phi /dF and x} with Eq. \eqref{eq: k from x}.
This ability to compute transition rates from combinations of forces and observables directly generalizes traditional Max Cal inference,\cite{jaynes_minimum_1980,evans_rules_2004,presse_principles_2013,davis_probabilistic_2018} which was limited to observable constraints. \\

The universal response laws extend to mixed coordinates too.
Maxwell-Onsager Relations directly follow from Eq. \eqref{eqs: computing x and F from d phi /dF and x}. And, the Fluctuation-Susceptibility Equalities can also be derived for the first time.  The response of the conjugate variables $(\textbf{x}_1,\boldsymbol{\mathfrak{F}}_2)$ to perturbations in the mixed coordinates $\boldsymbol{z}$ are given by the covariances (Method E): 
\begin{align} \label{eq: FSE in mixed coordinate}
\left[\begin{array}{cc}
\frac{\partial\textbf{x}_{1}}{\partial\boldsymbol{\mathfrak{F}}_{1}} & \frac{\partial\textbf{x}_{1}}{\partial\textbf{x}_{2}} \\ \\
\frac{\partial\boldsymbol{\mathfrak{F}}_{2}}{\partial\boldsymbol{\mathfrak{F}}_{1}} & \frac{\partial\boldsymbol{\mathfrak{F}}_{2}}{\partial\textbf{x}_{2}}
\end{array}\right]
&=\left[\begin{array}{cc}
\textbf{C}_{11} & \textbf{C}_{12}\\ \\
\textbf{0} & \textbf{I}
\end{array}\right]\left[\begin{array}{cc}
\textbf{I} & \textbf{0}\\ \\
-\textbf{C}_{22}^{-1}\textbf{C}_{21} & \textbf{C}_{22}^{-1}
\end{array}\right].
\end{align}
Here, $\textbf{C}_{11}$ is a $N_1$-by-$N_1$ covariance matrix of the components of $\textbf{X}_1$; $\textbf{C}_{22}$ is that of $\textbf{X}_2$; $\textbf{C}_{12}$ is a $N_1$-by-$N_2$ matrix with covariance terms between $\textbf{X}_1$ and $\textbf{X}_2$; And, $\textbf{C}_{21} = \textbf{C}_{12}^{\mathsf{T}}$ is the transpose of $\textbf{C}_{12}$.

\subsection*{Applications of the Two-Law Framework}
With its conjugated coordinates and potentials, CFT provides a principled way to model and analyze complex network flows. The power of this framework is best illustrated by its application to three distinct physical problems: optimizing a molecular machine, discovering a new transport law, and resolving a dynamic paradox.\\

\textbf{Constrained Optimization of Molecular Machines:}
Molecular machines are NEQ systems operating under constraints of forces, consuming a fixed source of chemical energy to produce directed motion. A key question is what design principles allow them to optimize performance, such as maximizing speed\cite{wagoner_mechanisms_2019,brown_theory_2020} or minimizing fluctuation.\cite{barato_thermodynamic_2015,horowitz_thermodynamic_2020} CFT provides the natural force-flux coordinates for this. A fixed chemical energy source (e.g., a constant ATP/ADP ratio) corresponds to a fixed cycle force, $\mathfrak{F}_{\text{cycle}}$, under the assumption of Local Detailed Balance.\cite{van_den_broeck_ensemble_2015,seifert_stochastic_2018,peliti_stochastic_2021,pachter_entropy_2024} The motor's performance can then be optimized by tuning the remaining degrees of freedom, which CFT identifies, such as the traffic on the chemical and mechanical cycle steps ($\tau_c, \tau_m$).\\

\begin{figure}
\begin{centering}
\includegraphics[width=\columnwidth]{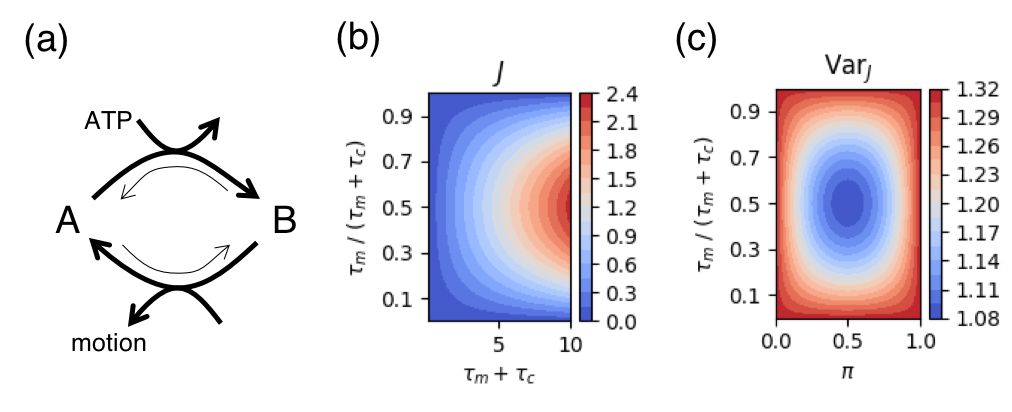}
\par\end{centering}
\caption{\textbf{Optimizing speed and minimizing fluctuation of a molecular motor under a fixed energy constraint.} (a) A two-state model of a molecular motor driven by chemical energy. (b) For a fixed cycle force ($\mathfrak{F}_{\text{cycle}}$), the motor's speed ($J$) is maximized when traffic is equally distributed between the chemical and mechanical steps. (c) The analysis of fluctuations ($\text{Var}_J$) shows that this ``equal-traffic principle'' also minimizes fluctuation for this model, even under additional physical constraints such as a fixed state population ($\pi$).
\label{fig: varJ}}
\end{figure}

Our framework reveals a new design principle for optimizing the speed of molecular motors. For a general unicircular process, such as the two-state model shown in (Fig. \ref{fig: varJ}a),\cite{wagoner_mechanisms_2019} we find an ``equal-traffic principle'' (derived in {Method F}): At a fixed cycle force and given total traffic (sum of all edge traffics), the cycle flux $J$ (the motor's speed) is maximized when the traffic is distributed equally among all edges in the cycle, illustrated in {(Fig. \ref{fig: varJ}b)}. This holds regardless of the state populations ($\pi$), revealing a degenerate optimum, a structure not straightforward to find before. Further, CFT allows for optimization under multiple constraints, showing how fluctuations in this model can be minimized by equal traffic even when the motor is constrained to operate with highly asymmetric state populations ($\pi\approx 1$) {(Fig. \ref{fig: varJ}c)}, a feature needed for motors like myosin.\cite{alberts_molecular_2002,piazzesi_skeletal_2007} \\

\textbf{A Third Kirchhoff's Law for Stochastic Transport:}
Kirchhoff’s classic laws for electrical circuits describe the conservation of current at nodes and voltage in loops, but they are macroscopic laws that neglect the underlying stochastic back-stepping of the carriers. For stochastic transport networks, such as particles diffusing through parallel channels, these fluctuations are significant. Here, we derive a ``third Kirchhoff's Law'' for stochastic transport. By treating edge traffic and its conjugate force as fundamental coordinates, CFT provides a new rule that governs network flows where fluctuations are significant.\\

\begin{figure}
\begin{centering}
\includegraphics[width= 0.9\columnwidth]{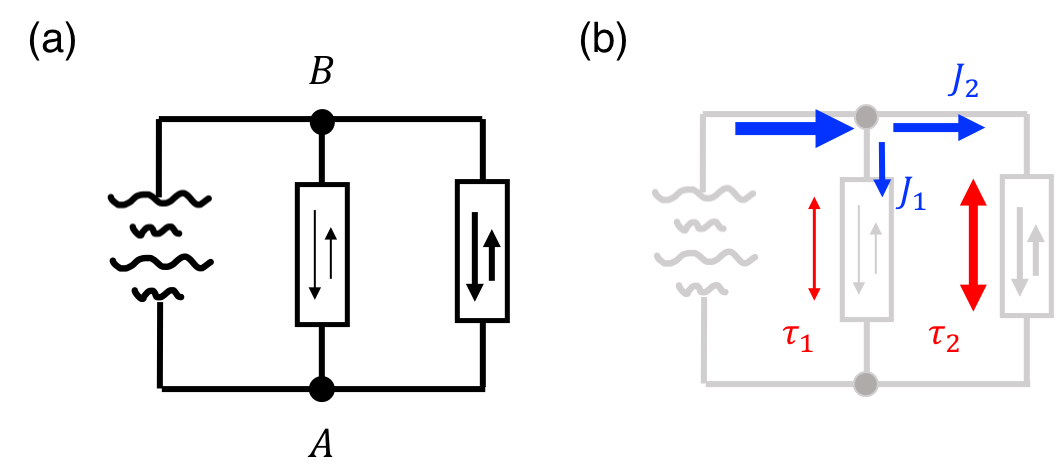}
\par\end{centering}
\caption{ \textbf{Kirchhoff's Third Law for Stochastic Flows}: {(a)} Stochastic flows on a circuit with a noisy source (wiggly battery icon) and asymmetric channels in parallel. (b) The framework reveals a new rule for symmetrically driven parallel channels: the ratio of net fluxes ($J_1/J_2$) is identical to the ratio of the underlying traffic ($\tau_1/\tau_2$), with both governed by the exponential of the edge forces.
 \label{fig: mesocircuit}}
\end{figure}

Consider a system where particles can traverse between two points A and B via multiple parallel channels {(Fig. \ref{fig: mesocircuit}a)}. When the two ``battery-channel cycles'' are driven out of equilibrium by equal cycle forces, corresponding to an equal level of asymmetry in the forward and reverse rates ($f_1/r_1=f_2/r_2$), CFT provides a simple and elegant rule. The ratio of net fluxes ($J_1/J_2$) through the two channels is identical to the ratio of their traffics ($\tau_1/\tau_2$), and both are determined by the exponential of their respective edge forces (Method G):
\begin{equation}
\frac{J_{1}}{J_{2}}=\frac{\tau_{1}}{\tau_{2}}=\frac{e^{\mathfrak{F}_{\text{edge},1}}}{e^{\mathfrak{F}_{\text{edge},2}}}=\frac{\sqrt{f_{1}r_{1}}}{\sqrt{f_{2}r_{2}}} 
\end{equation}
This result establishes the edge force as the fundamental quantity governing both the allocation of directed flow and the underlying stochastic activity in symmetrically driven networks, adding a new principle to transport laws where fluctuations are important and inviting future work on more complex networks.\\

\begin{figure}
\begin{centering}
\includegraphics[width= 0.9\columnwidth]{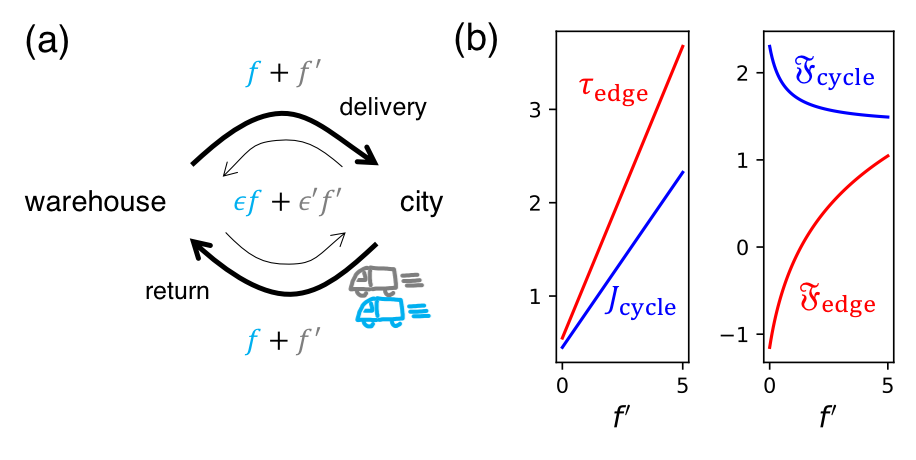}
\par\end{centering}
\caption{\textbf{Resolving a dynamic paradox with a complete force basis.} (a) A model of two drivers contributing to a delivery service. (b) As the driver with higher error rate delivers more (rate $f'$), the team's net delivery flux ($J_{\text{cycle}}$) increases, but the dynamic conjugate driving force ($\mathfrak{F}_{\text{cycle}}$) paradoxically decreases. This apparent contradiction is resolved by our complete coordinate system, which shows that the positive contribution from the simultaneously increasing edge force ($\mathfrak{F}_{\text{edge}}$) dominates.\label{fig: non-mono truck}}
\end{figure}

\textbf{The Necessity of a Complete Description:}
A key feature of CFT is the completeness of its coordinate system. Neglecting any of the three components---e.g. focus narrowly on $\pi,J,$ and $\mathfrak{F}_{\text{cycle}}$---can lead to an incomplete physical picture and apparent ``paradoxes.'' We demonstrate this with a simple, non-thermal model where two truck drivers, Alice and Bob, contribute to a delivery service {(Fig. \ref{fig: non-mono truck}a)}. If Bob, who has a higher error rate  than Alice's ($\epsilon'>\epsilon$), increases his forward delivery rate ($f'$), the team’s overall net flux of deliveries ($J_{\text{cycle}}$) increases, as expected. ``Paradoxically'', the conjugate cycle force ($\mathfrak{F}_{\text{cycle}}$), often considered the sole driver of the flux, simultaneously decreases. {(Fig. \ref{fig: non-mono truck}b)}\\

This anti-correlation between a flux and its putative driving force can not be explained if one only considers cycle forces and fluxes. The paradox is resolved within CFT by recognizing the crucial role the edge force, $\mathfrak{F}_{\text{edge}}$. As Bob works faster, the total edge force increases too. A full decomposition of the response in the force coordinate shows that the positive contribution from the increasing edge force overwhelms the negative contribution from the decreasing cycle force ({Method H}): \begin{equation}
\frac{{\rm d}J_{\text{cycle}}}{{\rm d}f'}=2 \underset{>0}{\underbrace{\left(\frac{\partial J_{\text{cycle}}}{\partial\mathfrak{F}_{{\text{edge}}}}\right)}}\underset{>0}{\underbrace{\frac{\partial\mathfrak{F}_{{\text{edge}}}}{\partial f'}}}+\underset{>0}{\underbrace{\left(\frac{\partial J_{\text{cycle}}}{\partial\mathfrak{F}_{{\text{cycle}}}}\right)}}\underset{<0}{\underbrace{\frac{\partial\mathfrak{F}_{{\text{cycle}}}}{\partial f'}}}.
\end{equation}
This example shows that a complete dynamical description is necessary to understand network behavior, highlighting the importance of CFT's First Law of completeness.

\subsection*{A Synthesis of Nonequilibrium Frameworks}
CFT builds upon foundational pillars\cite{pachter_entropy_2024}---like Maximum Caliber (Max Cal), Large Deviation Theory (LDT), and Stochastic Thermodynamics (ST)---to create a complete and constructive framework with a true Legendre-conjugate structure. We discuss below how this synthesis extends the capabilities of each parent framework.\\ 

First, \textbf{CFT elevates Max Cal from a powerful inference tool to a complete, constructive theory}. Traditional Max Cal infers a unique process from a limited set of observable constraints, ensuring axiomatic consistency\cite{shore_axiomatic_1980,presse_nonadditive_2013} and equivalency to Bayesian conditioning.\cite{csiszar_conditional_1987,chetrite_nonequilibrium_2015,chetrite_variational_2015,yang_statistical_2023} Our framework reveals the hidden assumption of the models it infers: in constraining only a subset of observables, it implicitly fixes the forces conjugate to all unconstrained parts of the system (Sec. III of the SI). By pairing Max Cal's variational principle with a First Law of completeness, CFT reveals and removes this limitation. It provides the machinery to build models from any valid mixed coordinate set—where a subset of forces is constrained alongside a complementary set of observables—thereby establishing a more general and physically robust foundation.\\

Second, \textbf{CFT achieves for dynamics what is well-known for static probabilities in equilibrium statistical physics}, where a single log-partition function (the free energy) describes the entire family of canonical ensembles. CFT provides the dynamical equivalent by reframing the tools of LDT.\cite{touchette_large_2009,barato_formal_2015, bertini_macroscopic_2015,chetrite_nonequilibrium_2015} Where LDT uses abstract tilting parameters as auxiliary mathematical tools to probe the fluctuations of a single process, CFT's First Law provides the proper coordinates to elevate the parameters into fundamental forces, $\boldsymbol{\mathfrak{F}}$, that have clear physical meanings as affinities and parameterize the entire space of dynamics. This reveals the Caliber as the true ``free energy-like'' potential for nonequilibria, with the scaled-cumulant generating function of each specific process generated from the universal Caliber (Method D). CFT thus transforms the analytical tools of LDT into a concrete and generative physical framework.\\

Finally, \textbf{CFT builds upon ST by providing the complete Legendre-conjugate structure}. ST offers deep insights by connecting dynamics to energetics, through assuming Local Detailed Balance and computing the Entropy Production Rate (EPR) as a time irreversibility measure.\cite{schnakenberg_network_1976,hill_free_1989,van_den_broeck_ensemble_2015,ge_physical_2010,seifert_stochastic_2018,peliti_stochastic_2021} However, the EPR fails as a general variational potential,\cite{martyushev_restrictions_2014,vellela_stochastic_2008} as its derivative yields the conjugate forces of fluxes only in the linear-response regime ({Method I}). Furthermore, it was not clear what are the right forces that are Legendre conjugated to net fluxes: Hill considered cycle affinities;\cite{hill_free_1989} Schnakenberg called both the logarithm of edge flux ratio, $\ln(\pi_i k_{ij}/\pi_j k_{ji})$, and the cycle affinities forces;\cite{schnakenberg_network_1976} others called edge affinities, $\ln(k_{ij}/k_{ji})$, forces.\cite{owen_universal_2020,aslyamov_nonequilibrium_2024} CFT provides the two missing pieces: a complete coordinate and the conjugated potentials. CFT \textit{unambiguously} identifies (half of) the cycle affinity as the \textit{sole} conjugate force to the cycle flux, establishing the Legendre structure arbitrarily far from equilibrium.

\subsection*{Outlook}

Equilibrium physics (EQP) has long stood as principled, powerful and versatile.  And although EQP is limited by nearness to equilibria, it's two-laws structure has long been seen as the key to an ultimately more general framework for arbitrary non-equilibria.  Here, we achieve this in CFT for network flows by switching the conceptual framework: (i) from state entropies to path entropies, and (ii) from complete observable sets such as $(U, V, N)$ to complete observable sets of distributions and fluxes $(\pi, \tau, J)$.  This now unlocks the same mathematical power as EQP has---\textit{i.e.}, of Legendre transforms and Maxwell-like relations and fluctuation-response relations for general conjugate variable pairs, now for forces and flows.  A virtue of CFT---stemming from its foundation in entropy as envisioned by pioneers, like Szil\'{a}rd and Jaynes and others\cite{szilard_uber_1929, mandelbrot_derivation_1964, jaynes_information_1957, callen_thermodynamics_1985}---is its ability to go beyond thermal processes, like heat and work and particle flows, to also model the transport of any mobile agent. 

\section*{Acknowledgments}
We thank Charles Kocher, Olga Movilla Miangolarra, Jonathan Pachter, Rostam Razban, Yuhai Tu, Lakshmanji Verma, and Jin Wang for insightful feedback. Y.-J. expresses his deepest gratitude to Hong Qian for guiding him in learning the essential theoretical pieces used in this work. We are grateful for the financial support from the Laufer Center for Physical and Quantitative Biology at Stony Brook, the John Templeton Foundation  (Grant ID 62564), and NIH (Grant RM1-GM135136).


%

\clearpage
\section*{Methods}
We provide key framework details, derivations, and outline of calculations here. Calculation details are deferred to the Supplemental Information (SI) for concision.\\

We first remark that, while we focus on continuous-time Markov jump processes in the present work, the ``Two-Law'' principles can be extended to other Markov or non-Markov dynamics. See {Sec. I of the SI}. 

\renewcommand{\thefigure}{M\arabic{figure}}
\setcounter{figure}{0} 

\subsection*{A. Justification for Steady-State and Long-Term Limit \label{M-A: SS justification}}

Our framework is formulated for nonequilibrium steady states in the long-term limit. The necessity of such focus for a theory of processes has been discussed by H. Qian and one of us.\cite{yang_statistical_2023} Here are the concepts. First, the steady-state analysis allows us to \textit{isolate a system's dynamical rules} (i.e., its transition rates) from the transient effects of its initial conditions. Transient behaviors can be analyzed afterward by its underlying steady-state characteristics---an ideal parallel to the bifurcation analysis in the mathematics of dynamical systems.\cite{yang_potentials_2021} Time-inhomogeneous processes can also be analyzed by the ``frozen'' steady-state properties at each instance.\\

Second, the 
long-term limit ($L\rightarrow \infty$) is necessary for characterizing a process. Unlike a static system, such as a collection of spatially distributed units, a process is described by a dynamical rule (the ``generator,'' \textit{e.g.} the set of transition rates) that produces a trajectory of states over time, \textit{in principle indefinitely}. To capture the properties of the process itself, its statistics must be analyzed over (arbitrarily long) paths, not just vectors of states with a fixed length. The necessity of this long-trajectory limit naturally leads to using the path entropy rate as the core thermodynamic-like potential. These two concepts are intrinsically linked: in the long-term limit, empirical observables converge to their steady-state averages, making the analysis of the path entropy rate the natural way to study a system's inherent dynamics. Generalizations is needed to handle ``sweeping processes'' without a steady state.\cite{lasota_chaos_1994}

\subsection*{B. Constructing the Observable Basis\label{M-B: observable_basis}}

The observable coordinate $\textbf{x} = ({\pi}_{\text{node}}, {\tau}_{\text{edge}}, {J}_{\text{cycle}})$ is constructed to be complete and non-degenerate. While constructing the node probabilities ${\pi}_{\text{node}}$ and edge traffics ${\tau}_{\text{edge}}$ are straightforward, the basis for net fluxes, ${J}_{\text{cycle}}$, requires a specific construction to ensure a linearly-independent basis. This method is well-established,\cite{schnakenberg_network_1976,hill_free_1989} so we simply illustrate the two main steps with an example.\\

\begin{figure}
\begin{centering}
\includegraphics[width=0.9\columnwidth]{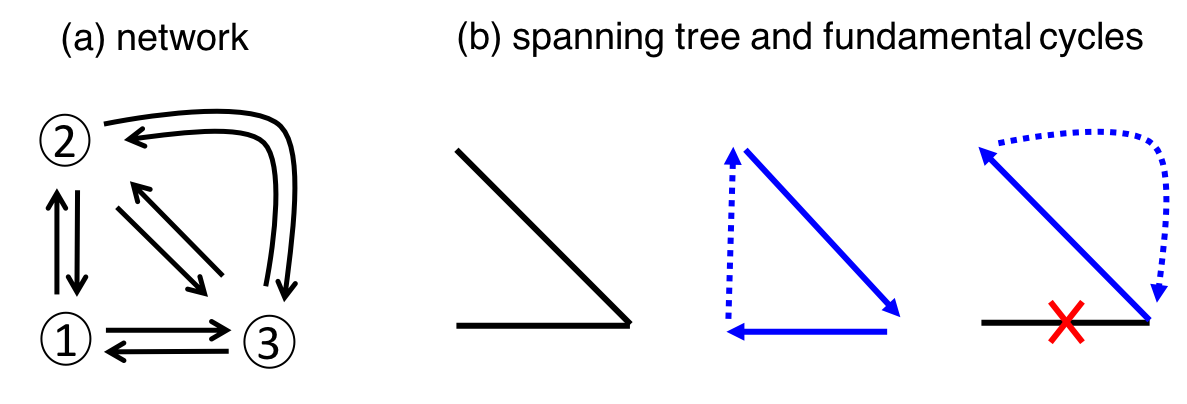}
\par\end{centering}
\caption{{\textbf{Steps to Find Fundamental Cycles}:} (a) An example network with multiple cycles. (b) A spanning tree of the network. Adding edges (dashed) forms the fundamental cycles. An edge on the spanning tree not used in a cycle is marked with a red x. \label{fig: tree-and-cycle-example}}
\end{figure}

First, a \textbf{spanning tree} is defined, which is a subgraph connecting all nodes without forming a cycle (Fig. \ref{fig: tree-and-cycle-example}b, left). The choice of spanning tree is not unique and can be engineered to best suit a system at hand.\\

Second, the remaining edges (the ``chords'') are added back one by one to define the fundamental cycles. Each one identifies a \textbf{fundamental cycle}, and redundant edges are removed (e.g. the red cross in  \ref{fig: tree-and-cycle-example}b). Adding the edge (1,2) back to the tree creates the fundamental cycle $c_1=1231$; Adding the curved edge (2,3) creates the cycle $c_2=232$.  The basis vector for cycle fluxes, $J_{\text{cycle}}$, is then composed of the net fluxes on these defining chord edges: $J_{c_1}=J_{12}$
  and $J_{c_2}=J^{\text{curv}}_{23}$.
All other net fluxes on the spanning tree can be written as a linear superposition of this basis. For instance, the net flux on the diagonal tree edge (2,3) is given by the superposition $J_{23}^{\text{diag}} =J_{c_1}-J_{c_2}$.\\

The rate observables in the coordinate $\textbf{x}$ are the long-term averages of corresponding extensive path observables, $\textbf{X}(\omega_L)$. For a length-$L$ trajectory $\omega_L$ with state $i_t$ at time $t$, these counting variables are defined by combinations of Kronecker delta functions $\delta_{i,j}$:\begin{subequations} \label{eqs: counting variables}
    \begin{align}
        X_{\text{node},n} &= \int_0^L \delta_{i_t,n} ~\text{d}t,\label{eq: Xnode}\\
        X_{\text{edge},ij} &= \sum_{\text{jump } l} \left(\delta_{i_{t_l^-},i} ~\delta_{i_{t_l^+},j} + \delta_{i_{t_l^-},j} ~\delta_{i_{t_l^+},i}\right),\label{eq: Xedge}\\
        X_{\text{cycle},c} &= \sum_{\text{jump }l} \left(\delta_{i_{t_l^-},a} ~\delta_{i_{t_l^+},b} - \delta_{i_{t_l^-},b} ~\delta_{i_{t_l^+},a}\right)\label{eq: Xcycle}
    \end{align}
\end{subequations}
where $t_l^\pm$ is the time before and after the jump time $t_l$ and $(a,b)$ is the defining edge of a fundamental cycle $c$. Taking the long-term limit of these counting variables ($\lim_{L\rightarrow \infty} X/L$) yields the corresponding rate observables $(\pi_{\text{node},n},\tau_{\text{edge},ij},J_{\text{cycle},c}).$

\subsection*{C. Deriving the Force Expressions}
While the force expressions can be derived by directly differentiating the path entropy rate, a more general proof comes from the Max Cal/LDT framework. This framework provides a formula for the posterior transition rates, $k_{ij}$, that result from maximizing the path entropy rate subject to constraints on the average rates of a set of path observables $\textbf{B}(\omega_L) = \int_0^L \textbf{f}(i_t) dt + \sum_{t_l} \textbf{g}(i_{t_l^-},i_{t_l^+})$.
The Max Cal posterior rates take the following form:\cite{mullins_analysis_1959,csiszar_conditional_1987,barato_formal_2015,chetrite_nonequilibrium_2015}
\begin{equation} \label{eq: posterior rate}
k_{ij}=\begin{cases}
k_{ij}^{(0)}\frac{r_{j}}{r_{i}}e^{\boldsymbol{\mathfrak{F}} \cdot \textbf{g}(i,j)} & ,i\neq j\\
k^{(0)}_{ii} + \boldsymbol{\mathfrak{F}} \cdot \textbf{f}(i) - \mathfrak{c} & ,i=j,
\end{cases}
\end{equation}
where $\boldsymbol{\mathfrak{F}}$ are the forces (Lagrange multipliers), $r_x(\boldsymbol{\mathfrak{F}})$ and $\mathfrak{c}(\boldsymbol{\mathfrak{F}})$ are the right eigenvector and largest eigenvalue of ``the tilted matrix''. See {Sec. II.B of the SI} for a brief review.\\

Our derivation applies this general formula to the specific, complete path observables $\textbf{X}$ in Eqs. \eqref{eqs: counting variables}. This is done by specifying the functions $\textbf{f}$ and $\textbf{g}$ as the appropriate Kronecker delta functions that select for node dwellings, edge traffics, and cycle fluxes. We set the prior to the doldrum process ($k_{ij}^{(0)}=1$) and solve for the forces $\boldsymbol{\mathfrak{F}}$ in terms of the rates $k_{xy}$.\\

\textbf{Node and Edge Forces:}
For the diagonal elements, we have two cases---the nodes $n \neq m$ (where $m$ is the reference node) and the node $m$:
\begin{align}
    k_{nn} + N_n &= \mathfrak{F}_n - \mathfrak{c}(\boldsymbol{\mathfrak{F}}) \\
    k_{mm} + N_m &= -\mathfrak{c}(\boldsymbol{\mathfrak{F}})
\end{align}
where $N_n$ is the number of edges connected to the node $n$.
Subtracting the two equations eliminates the Caliber $\mathfrak{c}$ and yields the node force:
\begin{equation}
    \mathfrak{F}_{\text{node}, n} = (k_{nn}+N_n) - (k_{mm}+N_m),
\end{equation}
which becomes the main text form via $k_{nn}=-\sum_{j(\neq n)}k_{nj}$ and $N_n = \sum_{j(\neq n)}1.$
For the off-diagonal elements, the multiplication of $k_{ij}$ and $k_{ji}$ cancels both the $r$ factor and the cycle force part due to symmetry, giving us the edge forces:
\begin{equation}
    \mathfrak{F}_{\text{edge}, ij} = \frac{1}{2}\ln(k_{ij}k_{ji}).
\end{equation}~\\

\textbf{Cycle Forces:}
To find the cycle forces, we take the logarithm of the off-diagonal relation for a transition $x \to y$:
\begin{equation}
    \frac{1}{2}\ln\frac{k_{xy}}{k_{yx}} = \ln\frac{r_y}{r_x} + \frac{1}{2}\boldsymbol{\mathfrak{F}} \cdot \left(\textbf{g}(x,y) - \textbf{g}(y,x)\right).
\end{equation}
The term $\boldsymbol{\mathfrak{F}} \cdot (\textbf{g}(x,y)-\textbf{g}(y,x))$ is non-zero only for the cycle forces, and only on their respective defining edges of the fundamental cycles. When we sum this expression around a fundamental cycle $c$ (a sequence of states $i_0 \to i_1 \to \dots \to i_\sigma \to i_0$), the ``gradient'' terms $\ln(r_y/r_x)$ telescope and cancel out. This isolates the force for that cycle:
\begin{equation}
    \mathfrak{F}_{\text{cycle}, c}  = \frac{1}{2}\ln \frac{k_{i_0 i_1} k_{i_1 i_2} \cdots k_{i_\sigma i_0}}{k_{i_1 i_0} k_{i_2 i_1}\cdots k_{i_0 i_\sigma}}.
\end{equation}
This concludes the derivation.

\subsection*{D. Caliber as the Universal Generating Function}
\label{appendix: Caliber and SCGF}

This section details how the Caliber potential, $\mathfrak{c}(\boldsymbol{\mathfrak{F}})$, is calculated and how it serves as a universal generating function for the system's fluctuation statistics.\\

\textbf{Calculating the Caliber via the Tilted Matrix:}
The Caliber is the largest eigenvalue of a ``tilted matrix,'' $\tilde{\textbf{M}}(\boldsymbol{\mathfrak{F}})$. Its construction is best understood by its connection to the posterior rate expression in Eq. \eqref{eq: posterior rate}. The off-diagonal elements, $\tilde{\textbf{M}}_{ij}$, correspond to the part of the posterior rate that modifies the prior, excluding the eigenvector ratio term: $\tilde{\textbf{M}}_{ij}=k^{(0)}_{ij} \exp(\boldsymbol{\mathfrak{F}}\cdot \textbf{g}(i,j))$ for $i\neq j$. The diagonal elements are similarly modified: $\tilde{\textbf{M}}_{ii}=k^{(0)}_{ii}+ \boldsymbol{\mathfrak{F}}\cdot \textbf{f}(i)$. We illustrate this using the 3-node network from Fig. \ref{fig: tree-and-cycle-example}a, which has two fundamental cycles: $c_1=1231$ and $c_2=232$. Setting node 1 as the reference ($m=1$) and using unit-rate doldrum state $k^{(0)}_{ij}=1$, the tilted matrix is: \begin{equation} 
\tilde{\textbf{M}}(\boldsymbol{\mathfrak{F}})=\left[\begin{array}{cccc}
-2 & e^{\mathfrak{F}_{12}+\mathfrak{F}_{c_1}} &  e^{\mathfrak{F}_{13}}\\
e^{\mathfrak{F}_{12}-\mathfrak{F}_{c_1}} & \mathfrak{F}_{2}-3 & e^{\mathfrak{F}_{23}} + e^{\mathfrak{F}_{23}'+\mathfrak{F}_{c_2}}\\
e^{\mathfrak{F}_{13}} &e^{\mathfrak{F}_{23}} + e^{\mathfrak{F}_{23}'-\mathfrak{F}_{c_2}} & \mathfrak{F}_{3}-3
\end{array}\right]\nonumber
\end{equation}
where $\mathfrak{F}_{23}$ is the edge forces on the diagonal 23 and $\mathfrak{F}'_{23}$ is that on the curved 23. The largest eigenvalue of this matrix is the Caliber, $\mathfrak{c}(\boldsymbol{\mathfrak{F}})$. See Sec. II.B of the SI for a derivation.\\

\textbf{Relation between the Caliber and the Scaled Cumulant Generating Functions:}
A key consequence of our ``Two-Law'' construction is that the entire phase space of dynamics is established as a single, unified exponential family. The Caliber serves as its log-partition function and thus as a universal generating function for cumulants. This is shown by relating the Caliber to the Scaled Cumulant Generating Function (SCGF):\cite{touchette_large_2009,barato_formal_2015}
\begin{equation}
g_{\boldsymbol{\mathfrak{F}}}(\boldsymbol{\theta}) = \lim_{L\to\infty} \frac{1}{L} \ln \sum_{\omega} P_{\boldsymbol{\mathfrak{F}}}(\omega) e^{\boldsymbol{\theta} \cdot \textbf{X}(\omega)}
\label{eq:scgf_def}
\end{equation}
where $\boldsymbol{\theta}$ are auxiliary variables (set to zero after derivative calculations) and $P_{\boldsymbol{\mathfrak{F}}}(\omega)$ is the path probability for a process with forces $\boldsymbol{\mathfrak{F}}$. Because every process is part of the same exponential family in CFT, its path probability takes the Boltzmann-like form derived from the Maximum Caliber principle ({Sec. II.A of the SI}):  $P_{\boldsymbol{\mathfrak{F}}}(\omega) = P_u(\omega) e^{\boldsymbol{\mathfrak{F}} \cdot \textbf{X}(\omega)} / \mathcal{Z}(\boldsymbol{\mathfrak{F}})$, where $\mathcal{Z}(\boldsymbol{\mathfrak{F}})$ is the dynamic partition function. Thus, we get:
\begin{equation}
    \sum_{\omega} P_{\boldsymbol{\mathfrak{F}}}(\omega) e^{\boldsymbol{\theta} \cdot \textbf{X}(\omega)}
    =\frac{\sum_{\omega} P_u(\omega) e^{(\boldsymbol{\mathfrak{F}}+\boldsymbol{\theta}) \cdot \textbf{X}(\omega)}}{\mathcal{Z}(\boldsymbol{\mathfrak{F}})} = \frac{\mathcal{Z}(\boldsymbol{\mathfrak{F}}+\boldsymbol{\theta})}{\mathcal{Z}(\boldsymbol{\mathfrak{F}})}.\nonumber
\end{equation}
With $\mathfrak{c}(\boldsymbol{\mathfrak{F}}) = \lim_{L\to\infty} (1/L) \ln \mathcal{Z}(\boldsymbol{\mathfrak{F}})$, we further get:
\begin{subequations}
\begin{align}
    g_{\boldsymbol{\mathfrak{F}}}(\boldsymbol{\theta}) &= \lim_{L\to\infty} \frac{1}{L} \left[ \ln \mathcal{Z}(\boldsymbol{\mathfrak{F}}+\boldsymbol{\theta}) - \ln \mathcal{Z}(\boldsymbol{\mathfrak{F}}) \right]\\
    &= \mathfrak{c}(\boldsymbol{\mathfrak{F}}+\boldsymbol{\theta}) - \mathfrak{c}(\boldsymbol{\mathfrak{F}}).
    \label{eq:caliber_scgf_relation}
\end{align}
\end{subequations}
This explicitly shows that the Caliber $\mathfrak{c}(\boldsymbol{\mathfrak{F}})$ serves as the universal generating function for the cumulants of the observables $\textbf{X}$ for any process under the force coordinates $\boldsymbol{\mathfrak{F}}$.

\subsection*{E. Mixed Coordinates via Partial Legendre Transform}
\label{appendix: derivatives in the mixed coordinate}

The thermodynamic-like potential $\phi(\boldsymbol{z})$ for a mixed coordinate $\boldsymbol{z}=(\boldsymbol{\mathfrak{F}}_1, \textbf{x}_2)$ is defined by a partial Legendre transform of either the path entropy or the Caliber (Eqs. \ref{eqs: phi} in the main text), and its derivatives yield the remaining conjugate variables. To show this, we differentiate $\phi(\boldsymbol{z})$ in its different forms.  Differentiating the form in Eq. \eqref{eq: phi and s} gives:
\begin{equation}
\frac{\partial \phi}{\partial \boldsymbol{\mathfrak{F}}_1} = \textbf{x}_1+\left(\boldsymbol{\mathfrak{F}}_1 + \frac{\partial \mathfrak{s}_{\text{path}}}{\partial \textbf{x}_1}\right)\cdot \frac{\partial \textbf{x}_1}{\partial \boldsymbol{\mathfrak{F}}_1} = \textbf{x}_1(\boldsymbol{z}) \label{eq:x1_from_phi}
\end{equation}
The term in parentheses vanishes because $\boldsymbol{\mathfrak{F}}_1 = -\partial\mathfrak{s}_{\text{path}}/\partial\textbf{x}_1$. Similarly, differentiating the form in Eq. \eqref{eq: phi and c} confirms that $\boldsymbol{\mathfrak{F}}_2(\boldsymbol{z})=-\partial\phi/\partial\textbf{x}_2$.\\

The fluctuation-susceptibility equalities in the mixed coordinate (Eq. \ref{eq: FSE in mixed coordinate}) are derived by taking partial derivatives of the following equations with respect to the components of $\boldsymbol{z}$:
\begin{subequations}
    \begin{align}
        \textbf{x}_1(\boldsymbol{z}) &=\frac{\partial \mathfrak{c}}{\partial \boldsymbol{\mathfrak{F}}_1} [\boldsymbol{\mathfrak{F}}_1,\boldsymbol{\mathfrak{F}}_2(\boldsymbol{z})]\\
        \textbf{x}_2 &=\frac{\partial \mathfrak{c}}{\partial \boldsymbol{\mathfrak{F}}_2} [\boldsymbol{\mathfrak{F}}_1,\boldsymbol{\mathfrak{F}}_2(\boldsymbol{z})].
    \end{align}
\end{subequations}
Using the notation $\textbf{C}_{ij} = \partial^2 \mathfrak{c}/\partial \boldsymbol{\mathfrak{F}}_i \partial \boldsymbol{\mathfrak{F}}_j$ for the covariance matrix blocks, the derivatives are:
\begin{subequations}
    \begin{align}
        \frac{\partial{\textbf{x}_1}}{\partial \textbf{x}_2} & = \textbf{C}_{12} \frac{\partial \boldsymbol{\mathfrak{F}_2}}{\partial \textbf{x}_2}~~;~~
        \frac{\partial{\textbf{x}_1}}{\partial \boldsymbol{\mathfrak{F}}_1}  = \textbf{C}_{11}+ \textbf{C}_{12} \frac{\partial \boldsymbol{\mathfrak{F}}_2}{\partial \boldsymbol{\mathfrak{F}}_1}\\
        \textbf{I} & = \textbf{C}_{22}\frac{\partial \boldsymbol{\mathfrak{F}_2}}{\partial \textbf{x}_2}~~;~~
        \textbf{0}  =  \textbf{C}_{21} +\textbf{C}_{22} \frac{\partial \boldsymbol{\mathfrak{F}}_2}{\partial \boldsymbol{\mathfrak{F}}_1}.
    \end{align}
\end{subequations}
The last two equations give the derivatives of $\boldsymbol{\mathfrak{F}}_2$.
Substituting them back into the first two equations gives the derivatives of $\textbf{x}_1$, which gives Eq. \eqref{eq: FSE in mixed coordinate} in the main text.

\subsection*{F. Molecular Motor Optimization Analysis}
\textbf{General ``Equal Traffic Principle'' for Unicircular Processes:}
We first prove the ``Equal Traffic Principle'' for unicircular processes under the coordinate $\boldsymbol{z}=(\pi,\tau,\mathfrak{F}_{\text{cycle}})$, obtained from the Legendre transform from the observable coordinate $\textbf{x}=(\pi,\tau,J)$: \begin{equation} \label{eq: phi and spath in method}
    \phi=J ~\mathfrak{F}_{\text{cycle}}+\mathfrak{s}_{\text{path}}.
\end{equation}
The dual relation connecting the two coordinates are \begin{equation} \label{eq: dual between J and F cycle under fixed pi tau}
    J=\frac{\partial \phi~~~~~~}{\partial \mathfrak{F}_{\text{cycle}}};~~\mathfrak{F}_{\text{cycle}}=-\frac{\partial \mathfrak{s}_{\text{path}}}{\partial J~~~~~},
\end{equation}
which can be derived by implicit differentiation of Eq. \eqref{eq: phi and spath in method} under the two coordinates. For unicircular processes, Eqs. \eqref{eq: dual between J and F cycle under fixed pi tau} simplify and lead to our ``equal traffic principle.''\\

Consider a generic cycle with $N$ edges. We denote $\tau_n$ as the edge traffic  on the $n$th edge, $\tau_{\text{tot}}$ as their sum and $\rho_n=\tau_n/\tau_{\text{tot}}$ as the traffic distribution. Then, we can solve $J(\boldsymbol{z})$ by simply inverting $\mathfrak{F}_{\text{cycle}}=-\partial \mathfrak{s}_{\text{path}}/\partial J$:
\begin{subequations}
\begin{align}
    \mathfrak{F_{\text{cycle}}} & = \frac{1}{2}\sum_{n=1}^N \ln \frac{\rho_n+(J/\tau_{\text{tot}})}{\rho_n-(J/\tau_{\text{tot}})} \label{eq: J in tau and Fc for N-edge cycle} \\ \Rightarrow J &= \frac{\tau_{\text{tot}}}{N} G_N (\{\rho_n \},\mathfrak{F}_{\text{cycle}})
\end{align}
\end{subequations}
where the function $G_N(\cdot,\cdot)$ can be analytically solved for $N=2,3$ and numerically for larger $N$.\\

We now prove $G_N\le \tanh{\frac{\mathfrak{F}_{\text{cycle}}}{N}}$, which saturates when $\rho_n=1/N$. First, Eq. \eqref{eq: J in tau and Fc for N-edge cycle} can be rewritten as \begin{equation}
    \mathfrak{F}_{\text{cycle}} = \sum_{n=1}^{N} \text{arccoth}\left(\frac{\tau_n}{J}\right). 
\end{equation}
Then, since $\text{coth}(\cdot)$ is convex for positive arguments, Jensen's inequality states that 
\begin{subequations}
\begin{align}
    \text{coth}\left(\frac{\mathfrak{F}_{\text{cycle}}}{N}\right) &= \text{coth} \left[ \frac{1}{N} \sum_{n=1}^{N} \text{arccoth}(\tau_n/J)\right] \\
    &\le \frac{1}{N} \sum_{n=1}^N \frac{\tau_n}{J} = \frac{\tau_{\text{tot}}}{NJ} 
\end{align}
\end{subequations}
This proves that $G_N\le \tanh{\frac{\mathfrak{F}_{\text{cycle}}}{N}}$, the equal traffic principle, and that the bound is saturated with equal traffic distribution but is degenerate because $\pi$ does not enter the bound. Main text uses the $N=2$ case where $G_2(\rho,\mathfrak{F}_{\text{cycle}}) = \text{coth}(\mathfrak{F}_{\text{cycle}})-\sqrt{...}/[2 \sinh (\mathfrak{F}_{\text{cycle}})]$ with $\sqrt{...}=\sqrt{2+8\rho(1-\rho)+2(1-2\rho)^2\cosh(2\mathfrak{F}_{\text{cycle}})}$.  We set $\mathfrak{F}_{\text{cycle}}=1$ when presenting Fig. \ref{fig: varJ}b.\\

\textbf{Fluctuation Analysis for the Two-State Model:}
We connect to the force coordinate to compute the asymptotic fluctuation of the cycle flux, $V_J$. For simplicity, denote $\boldsymbol{\mathfrak{F}}=(\mathfrak{F}_{\text{node},A},\mathfrak{F}_{\text{edge},c},\mathfrak{F}_{\text{edge},m},\mathfrak{F}_{\text{cycle}})$ as $\boldsymbol{\mathfrak{F}}=(\alpha,\beta_c,\beta_m,\gamma)$. The tilted matrix is then \begin{equation}
    \tilde{M}=\left[\begin{array}{cc}
\alpha-2 & e^{\beta_c}+e^{\beta_{m}-\gamma}\\
e^{\beta_c}+e^{\beta_{m}+\gamma} & -2
\end{array}\right]
\end{equation}
and the Caliber $\mathfrak{c}$ is its largest eigenvalue: \begin{equation} \label{eq: caliber of the MM}
    \mathfrak{c} (\alpha,\beta_c,\beta_m,\gamma)=\frac{\square-\left(4-\alpha\right)}{2}
\end{equation} where $\square= [\left(4-\alpha\right)^{2}-8\left(2-\alpha\right)+4(e^{\beta_{c}}+e^{\beta_{m}+\gamma})$  $\left(e^{\beta_{c}}+e^{\beta_{m}-\gamma}\right)]^{1/2}.$ Taking its first and second derivatives give $J(\boldsymbol{\mathfrak{F}})$ and $V_J(\boldsymbol{\mathfrak{F}}).$ Further, using the rate expressions of forces $\boldsymbol{\mathfrak{F}}(k_{ij})$ and then $k_{ij}[\textbf{x}(\boldsymbol{z})]$ show that $\square=f_m+f_c+r_m+r_c = J/[2\pi_A(1-\pi_A)G_2].$ Integrating these  shows ({Sec. IV.A of the SI}) \begin{equation}
    \frac{V_{J}}{J}=\text{coth}(\gamma)-4\pi_A\left(1-\pi_A\right)G(\rho,\gamma).
\end{equation}
For a fixed $J$ and cycle force $\mathfrak{F}_{\text{cycle}}=\gamma$, the fluctuation is at minimum when $\pi_A=\rho=1/2$ or at constrained minimum when $\rho=1/2$ if under $\pi_A$ constraint. We set $J=1,\mathfrak{F}_{\text{cycle}}=1$ when presenting Fig. \ref{fig: varJ}c.

\subsection*{G. Derivation of the ``Third Kirchhoff's Law''}

\begin{figure}
\begin{centering}
\includegraphics[width=\columnwidth]{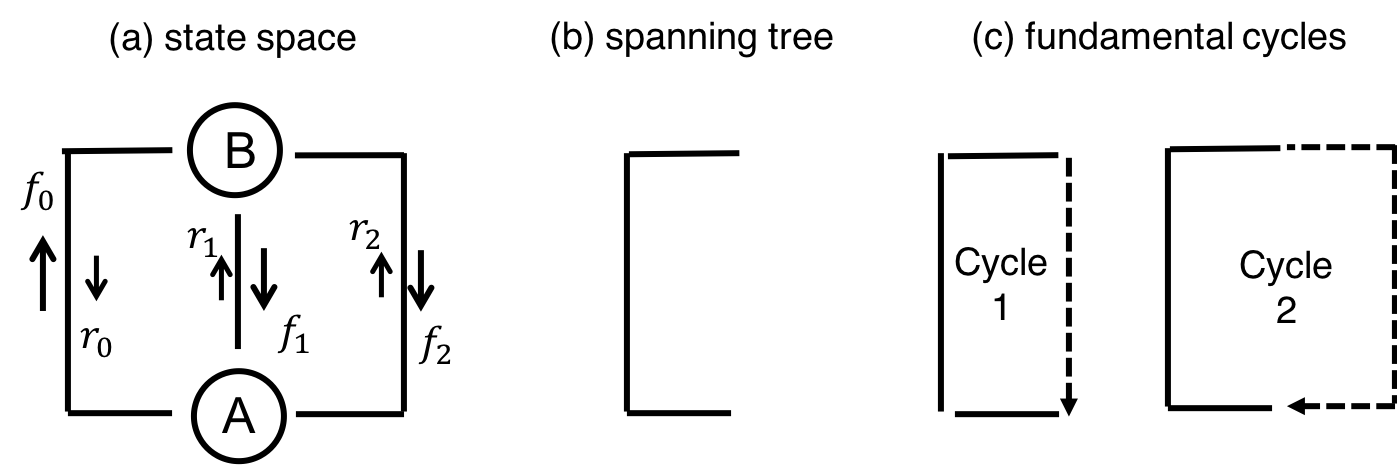}
\par\end{centering}
\caption{ \textbf{The cycle decomposition for the Kirchhoff example. } (a) shows the state space with two nodes and three edges. (b) is the spanning tree. (c) are the two fundamental cycles \label{fig: Kirchhoff cycle decomposition}}
\end{figure}

We consider the simple two-state-three-edge Markov dynamics for the Kirchhoff example of the main text. The dynamics is 6 dimensional with transition rates $f_i,r_i$, $i=0,1,2$ and state space illustrated in Fig. \ref{fig: Kirchhoff cycle decomposition}a. We choose the spanning tree in Fig. \ref{fig: Kirchhoff cycle decomposition}b to define the two fundamental cycles in Fig. \ref{fig: Kirchhoff cycle decomposition}c. To show ``the Kirchhoff's third law'' for stochastic transport. We use the force coordinate, which for simplicity we denote $\boldsymbol{\mathfrak{F}}=(\mathfrak{F}_{\text{node},A},\mathfrak{F}_{\text{edge},0},\mathfrak{F}_{\text{edge},1},\mathfrak{F}_{\text{edge},2},\mathfrak{F}_{\text{cycle},1},\mathfrak{F}_{\text{cycle},2})=(\alpha,\beta_0,\beta_1,\beta_2,\gamma_1,\gamma_2).$\\

The Caliber $\mathfrak{c}(\boldsymbol{\mathfrak{F}})=[\square'-(4-\alpha)]/2$ where $\square'= [\left(4-\alpha\right)^{2}-8\left(2-\alpha\right)+4(e^{\beta_{0}}+e^{\beta_{1}+\gamma_1}++e^{\beta_{2}+\gamma_2})$  $(e^{\beta_{0}}+e^{\beta_{1}-\gamma_1}++e^{\beta_{2}-\gamma_2})]^{1/2}$  is again the largest eigenvalue of the tilted matrix:
\begin{equation}
    \tilde{\textbf{M}}=\left[\begin{array}{cc}
\alpha-2 & e^{\beta_0}+e^{\beta_{1}-\gamma_{1}}+e^{\beta_{2}-\gamma_{2}}\\
e^{\beta_0}+e^{\beta_{1}+\gamma_{1}}+e^{\beta_{2}+\gamma_{2}} & -2
\end{array}\right].\nonumber
\end{equation}
The partial derivatives of the Caliber gives $\tau_1,\tau_2,J_1,$ and $J_2$. In general cases, one will get\begin{subequations}
    \begin{align}
         \frac{J_1}{J_2} &= \frac{e^{\beta_1}}{e^{\beta_{2}}} \cdot \frac{e^{\beta_0} \sinh(\gamma_{1})+e^{\beta_{2}} \sinh(\gamma_{1}-\gamma_{2})}{e^{\beta_{0}} \sinh(\gamma_{2})+e^{\beta_{1}} \sinh(\gamma_{2}-\gamma_{1})}\nonumber\\
        \frac{\tau_1}{\tau_2}&=\frac{e^{\beta_1}}{e^{\beta_2}}\cdot\frac{e^{\beta_0}\cosh(\gamma_1)+e^{\beta_2}\cosh(\gamma_1-\gamma_2)+e^{\beta_1}}{e^{\beta_0}\cosh(\gamma_2)+e^{\beta_1}\cosh(\gamma_2-\gamma_1)+e^{\beta_2}}. \nonumber
    \end{align}
\end{subequations}
When the cycle forces match $\gamma_1=\gamma_2$, these ratios simplify to the desired $J_1/J_2=\tau_1/\tau_2=e^{\beta_1}/e^{\beta_2}$. Since $\gamma_1=\ln \sqrt{(f_0f_1)/(r_0r_1)}$ and $\gamma_2=\ln \sqrt{(f_0f_2)/(r_0r_2)}$, equal cycle forces mean equal rate asymmetry, $\gamma_1=\gamma_2\Leftrightarrow f_1/r_1=f_2/r_2.$\\

By computing $\tau_0$, one can further show that, at EQ $(\gamma_1=\gamma_2=0)$,  traffic distribution is solely determined by the edge forces (Sec. IV.B of the SI), \begin{equation}
\tau_0:\tau_1:\tau_2=e^{\beta_0}:e^{\beta_1}:e^{\beta_2}.     
\end{equation}

\subsection*{H. Decomposition of Flux-Force Anti-Correlation}
The decomposition of responses into force components is a powerful analytical tool of CFT. Its utility is built on a fundamental difference: forces $\boldsymbol{\mathfrak{F}}$ are functions of local set of rates---depending only on rates attached to specific nodes, on the edge, or on the cycle---while observables $\textbf{x}$ are global because they are steady-state properties depending on all transition rates in the network. Therefore, tuning a local set of rates $\boldsymbol{k}$ often corresponds to changing a few but not all of the forces, allowing the system's global response to be cleanly decomposed in to a sum over elemental force responses: \begin{equation}
    \frac{\text{d} \textbf{x}}{\text{d} \boldsymbol{k}}=\sum_\mathfrak{F}\frac{\partial \textbf{x}}{\partial \mathfrak{F}}\cdot \frac{\partial \mathfrak{F}}{\partial\boldsymbol{k}}.
\end{equation} The susceptibility coefficients $\partial \textbf{x}/\partial \mathfrak{F}$ are given by the second derivatives of the Caliber and thus satisfy Fluctuation-Susceptibility and Maxwell-Onsager relations.\\

The Alice-Bob example applies this analysis. We assumed rotational symmetry for simplicity. This fixes the node forces and equate the contribution from the two edge forces, simplifying the change in cycle flux ($J_{\text{cycle}}$) with respect to the control parameter ($f'$) into:
\begin{equation}
    \frac{\mathrm{d}J_{\text{cycle}}}{\mathrm{d} f'} = 2 \left(\frac{\partial J_{\text{cycle}}}{\partial \mathfrak{F}_{\text{edge}}}\right) \frac{\partial \mathfrak{F}_{\text{edge}}}{\partial f'} + \left(\frac{\partial J_{\text{cycle}}}{\partial \mathfrak{F}_{\text{cycle}}}\right) \frac{\partial \mathfrak{F}_{\text{cycle}}}{\partial f'}. 
\end{equation} 
The derivatives of the forces with respect to $f'$ are calculated directly from their definitions:\begin{align*}
\frac{\partial\mathfrak{F}_{{\rm cycle}}}{\partial f'} & =-\frac{\left(\epsilon'-\epsilon\right)f}{\left(f+f'\right)\left(\epsilon f+\epsilon'f'\right)}<0;\\
\frac{\partial\mathfrak{F}_{{\rm edge}}}{\partial f'} & =\frac{1}{2}\left(\frac{1}{f+f'}+\frac{\epsilon'}{\epsilon f+\epsilon'f'}\right)>0.
\end{align*}  
The susceptibility terms, $\partial J_{\text{cycle}}/\partial \mathfrak{F}_{\text{cycle}}$ (the variance, which is positive) and $\partial J_{\text{cycle}}/\partial\mathfrak{F}_{\text{edge}}$ (the covariance), are calculated from the second derivatives of the Caliber, which is the same Caliber as the Molecular Motor example (Eq. \ref{eq: caliber of the MM}). A direct differentiation gives $J(\boldsymbol{\mathfrak{F}})$, and another derivative gives $\partial J_{\text{cycle}}/\partial\mathfrak{F}_{\text{edge}}=J/2>0$ (Sec. IV.C of the SI). For presenting Fig. \ref{fig: non-mono truck}b, we set $f=1, \epsilon=0.1,$ and $\epsilon'=0.25$.

\subsection*{I. Limitations of Entropy Production as a Potential}
Entropy production rate (EPR) is only a thermodynamic-like potential in the linear response regime:\cite{prigogine_time_1978} Its derivative only yields the conjugate force when fluxes are small, and it does not have a generally valid variational principle  for far-from-EQ systems.\cite{martyushev_restrictions_2014,vellela_stochastic_2008}  This can be shown with the two-state-two-edge Markov jump process in our molecular motor example. 
The cycle affinity is 
\begin{equation}
\mathcal{A}=\ln\frac{f_c f_m}{r_c r_m}=\ln \frac{(\tau_{c}+J)(\tau_{c}+J)}{(\tau_{m}-J)(\tau_{m}-J)}
\end{equation}
where $J$ is the net cycle flux and $\tau_c$ and $\tau_m$ are the edge traffic terms. The entropy production rate ($k_{\text{B}} T =1$ for simplicity) is $\text{EPR}=J\cdot \mathcal{A}$, which can be parameterized  by $(\tau_{c},\tau_{m},J)$.
Its derivative gives
\begin{equation}
    \frac{\partial ~\text{EPR}}{\partial J} =\mathcal{A}+\frac{2J\left(\tau_{c}+\tau_{m}\right)}{\left(\tau_{c}^{2}-J^{2}\right)\left(\tau_{m}^{2}-J^{2}\right)}
\end{equation}
which only equals $\mathcal{A}$ in the near-equilibrium limit where $|J|\ll\min(\tau_{c},\tau_{m})$.  With CFT, we now understand why EPR fails as a variational potential: it only captures the irreversibility part $\mathfrak{F}_{\text{cycle}}\cdot J_{\text{cycle}}$ of the full (negative) path entropy $(-\mathfrak{s}_{\text{path}})=\mathfrak{F}_{\text{node}}\cdot \pi_{\text{node}}+\mathfrak{F}_{\text{edge}}\cdot \tau_{\text{edge}}+\mathfrak{F}_{\text{cycle}}\cdot J_{\text{cycle}}-\mathfrak{c}.$


%

\end{document}